\documentclass{article}
\usepackage{spconf,amsmath,graphicx}

\usepackage{enumitem}
\usepackage{amssymb} 
\setlist{nosep, leftmargin=14pt}

\usepackage{mwe} % to get dummy images

\newcommand{\V}[1]{\textbf{#1}}
\title{Weakly Supervised Segmentation and Classification of Alpha-Synuclein Aggregates in Brightfield Midbrain Images}
\name{
\begin{tabular}{c}
Erwan Dereure$^{1}$ \qquad
Robin Louiset$^{2,3,4,5}$ \qquad
Laura Parkkinen$^{6}$ \\
David A Menassa$^{6,7,*}$\thanks{*~These authors contributed equally to this work.} \qquad
David Holcman$^{1,8,*}$\footnotemark[1] \qquad
\end{tabular}
\thanks{Corresponding author: dereure@bio.ens.psl.eu}
}
\address{$^{1}$\small{Group of Applied Mathematics and Computational Biology, Ecole Normale Sup\'erieure, PSL University, Paris, France.} \\$^{2}$\small{AP-HP, Hôpital Henri Mondor-Albert Chenevier, Service de Neurologie, F-94010 Créteil, France.}
\\$^{3}$\small{INSERM U955, Institut Mondor de Recherche Biomédicale, UPEC, Equipe NeuroPsychologie Interventionnelle, F-94010 Creteil, France.} 
\\$^{4}$\small{Département d'Etudes Cognitives, École normale supérieure, PSL University, 75005 Paris, France.} 
\\$^{5}$\small{NeurATRIS, Créteil, France.} \\$^{6}$\small{Nuffield Department of Clinical Neurosciences and the Queen’s College, University of Oxford, UK.}\\$^{7}$\small{Department of Women’s and Children’s Health, Karolinska Institutet, Sweden.}\\$^{8}$\small{Churchill College, Cambridge University, CB30DS UK.}}
\usepackage{eso-pic}
\newlength{\xoffset}
\newlength{\xoffsetbis}
\newlength{\yoffset}
\newlength{\yoffsetbis}
\AddToShipoutPicture{
        \AtPageLowerLeft{\put(\LenToUnit{\xoffset},\LenToUnit{\yoffset}){
\rotatebox[origin=r]{-90}{\color{gray}{\Large{This work has been submitted to the IEEE for possible publication.}}}}}
        \AtPageLowerLeft{\put(\LenToUnit{\xoffsetbis},\LenToUnit{\yoffsetbis}){
        \rotatebox[origin=r]{-90}{\color{gray}{\Large{Copyright may be transferred without notice, after which this version may no longer be accessible.}}}
}}}

\begin{document}
\ninept
\maketitle
%%%%%%%%%%%%%%%%%%%%%%%%%%%%%%%%%%%%%%%%%%%%%%%%%%%
\begin{abstract}
Parkinson’s disease (PD) is a neurodegenerative disorder associated with the accumulation of misfolded alpha-synuclein aggregates, forming Lewy bodies and neuritic shape used for pathology diagnostics. Automatic analysis of immunohistochemistry histopathological images with Deep Learning provides a promising tool for better understanding the spatial organization of these aggregates.\\
In this study, we develop an automated image processing pipeline to segment and classify these aggregates in whole-slide images (WSIs) of midbrain tissue from PD and incidental Lewy Body Disease (iLBD) cases based on weakly supervised segmentation, robust to immunohistochemical labelling variability, with a ResNet50 classifier. Our approach allows to differentiate between major aggregate morphologies, including Lewy bodies and neurites with a balanced accuracy of $80\%$. This framework paves the way for large-scale characterization of the spatial distribution and heterogeneity of alpha-synuclein aggregates in brightfield immunohistochemical tissue, and for investigating their poorly understood relationships with surrounding cells such as microglia and astrocytes. 
\end{abstract}
%%%%%%%%%%%%%%%%%%%%%%%%%%%%%%%%%%%%%%%%%%%%%%%%%%5%
\begin{keywords}
  Image processing, Segmentation, Parkinson's disease, Alpha-Synuclein, Image Retrieval, Shape Classification
\end{keywords}
%%%%%%%%%%%%%%%%%%%%%%%%%%%%%%%%%%%%%%%%%%%%%%%%%%
\section{Introduction}
\vspace{-1mm}

\begin{figure*}[!t]
\centerline{\includegraphics[width=1.95\columnwidth]{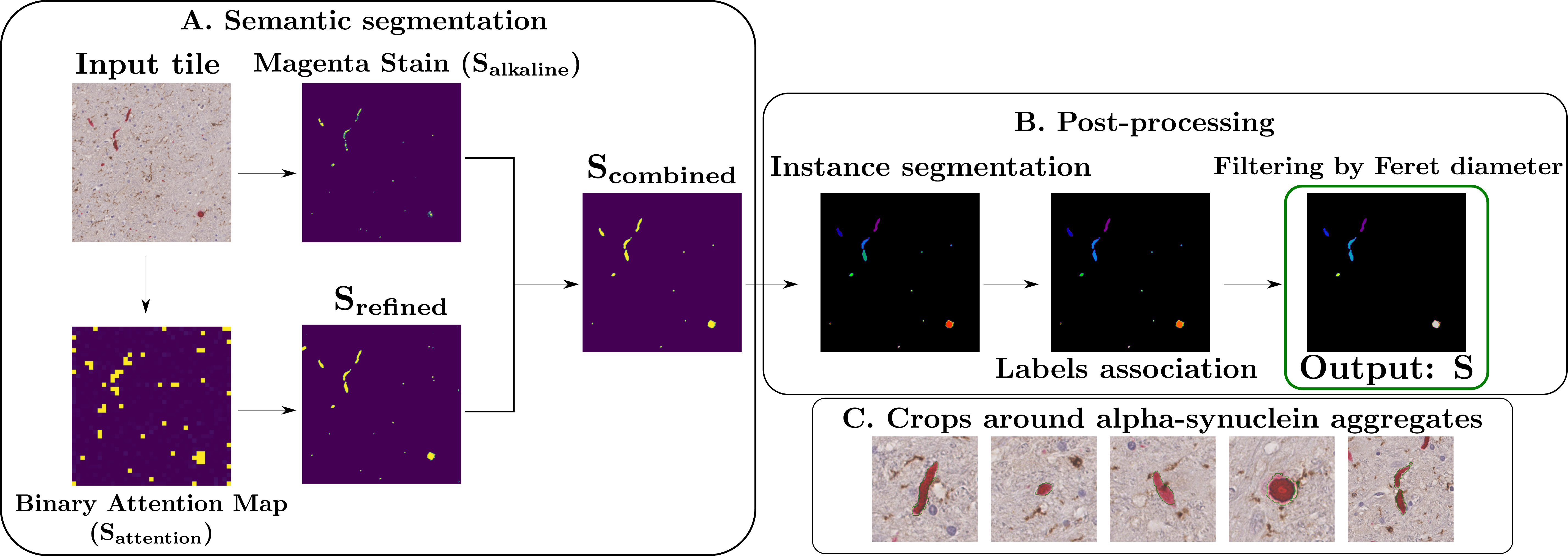}}
\caption{{\bf Overview of the segmentation pipeline for alpha-synuclein aggregates in brightfield immunohistochemistry.} {\bf (A)} Input tile undergoes semantic segmentation using a vision transformer classifier. The resulting attention map is refined with a fully connected Conditional Random Field (CRF) and combined with stain decomposition to isolate magenta-stained regions corresponding to alpha-synuclein. {\bf (B)} Instance segmentation is performed on the refined binary mask to extract individual aggregates, followed by post-processing steps including label association and small object filtering. Output: instance segmentation mask $\V{S}$ of alpha-synuclein aggregates. {\bf(C)} Final output: $256 \times 256$ image crops centered on each segmented alpha-synuclein aggregate, used for downstream classification, with segmentation contours in green.}
\label{fig:segmentation_pipeline}
\end{figure*}

%%%%%%%%%%%%%%%%%%%%%%%%%%%%%%%%%%%%%%%%%%%%%%%%%%
Parkinson’s disease (PD) is a hypokinetic disorder characterized by the inability to
generate voluntary movement \cite{dauer2003parkinson, cheng2010clinical}. The neurodegeneration of the nigrostriatal pathway underlies the symptoms and neuropathologically, dopaminergic neuronal cell loss of up to $80\%$ has been documented in the ventral substantia nigra pars compacta in patients in the late stages of the disease \cite{dauer2003parkinson}. Misfolded phosphorylated alpha-synuclein aggregates accumulate in neurons as Lewy bodies or in neuronal processes as Lewy neurites and are thought to cause neurodegeneration \cite{altay2022prominent}. These alpha-synuclein aggregates can also be present in incidental Lewy Body Disease (iLBD) \cite{dijkstra2014stage}, considered to represent early or prodromal stages of Parkinson’s disease \cite{prasad2012critical}. Concomitant with neurodegeneration is a state of sustained chronic neuroinflammation centrally orchestrated by microglia, macrophages and astrocytes with increasing evidence suggesting the involvement of the peripheral immune system \cite{roodveldt2024immune, andersen2025sympathetic}. \\
To study interactions between microglia and alpha-synuclein aggregates, our aim is to  quantify the spatial distribution, morphology, and organization of aggregates, to  ultimately reveal their count, fraction, shape heterogeneity, and associations with microglial phenotypes \cite{perochon2025unraveling}. Neuropathological evaluation of alpha-synuclein pathology has traditionally relied on manual or semi-quantitative approaches, with neuroinflammatory responses remaining poorly characterized. Typically, neuropathologists classify alpha-synuclein aggregates by their morphology. In this work, we present an automated pipeline designed to facilitate large-scale analysis.
% We propose an automated approach to detect and classify alpha-synuclein pathology phenotypes in the midbrain of PD and iLBD cases, combining image processing techniques and deep learning algorithms. 
% We further apply spatial statistical analyses, including level-set–based methods, to estimate coupling distances and association strengths between microglial phenotypes, neurons, and pathological features. 
% Finally, clustering algorithms are used to reveal the spatial organization of microglia around pathological aggregates.
% (!!! TO REWRITE !!!
% Here, we use a foundational large
% language model to automatically classify all features of alpha synuclein pathology
% and microglial phenotypes in the midbrain of PD and iLBD cases. This allows us to
% calculate distributions, numbers, proportions and to document the heterogeneity of
% shapes. We also apply spatial statistics including the levelset analyses to measure
% coupling distances and strengths of associations between pairs (microglia
% phenotype, neuron or pathological feature) and clustering algorithms to unravel how
% microglia are organised around pathological features. We validate our algorithms
% with expert neuropathologists and assess precision and sensitivity against a
% commercial platform. This tool is packaged into this and is available to adapt to your
% own images and will be useful for the community.)
The first stage of the pipeline involves segmenting alpha-synuclein aggregates, which are immunohistochemical labeled (hereafter referred to as stained for simplicity) in magenta in our images (Figure \ref{fig:segmentation_pipeline}). Few studies have addressed this type of segmentation: some analyses rely on channel intensity thresholding only \cite{vatsa2024network, pearce2022automatic}, this is not robust to impurities in staining, as shown in Figure \ref{fig:segmentation_pipeline}, where brown stain contamination of the magenta channel disrupts stain-based segmentation. Other studies employ machine learning–based tools such as Aiforia \cite{barber2025development} or Ilastik \cite{dadgar2022mesoscale}, which depend on weak annotations for object detection. Although these annotations are weak, their generation remains costly and time-consuming, as large quantities are required and the process lacks flexibility for more complex tasks, making it unsuitable for large-scale analysis. Then, to better characterize the tissue, the segmented aggregates must be further classified. Here our contribution is three-fold by  
\begin{itemize}  
    \item {\bf Designing a weakly supervised segmentation pipeline}: it is robust to staining impurities, which requires only tile-level binary annotations and leverages self-attention maps and stain matrix factorization to produce high-quality segmentation masks suitable for downstream quantitative analyses, including shape characterization and spatial statistics.
    \item {\bf Developing a self-supervised nearest-neighbors retrieval algorithm.} This algorithm could assist neuropathologists in generating annotation.  
    \item {\bf Using these annotations to train a neural network classifier:} The network will be capable of automatically distinguishing aggregate subgroups. Importantly, this classification does not rely on segmentation masks, as this network provides greater flexibility in differentiating between challenging aggregate types.
\end{itemize}  
\vspace{-1mm}
%%%%%%%%%%%%%%%%%%%%%%%%%%%%%%%%%%%%%%
\vspace{-1mm}
%%%%%%%%%%%%%%%%%%%%%%%%%%%%%%%%%%%%%%
\section{Data description}
\vspace{-2mm}
%%%%%%%%%%%%%%%%%%%%%%%%%%%%%%%%%%%%%%
Our dataset is composed of 44 Whole Slide Images (WSI) of size approximately $100 000 \times 100 000$ pixels. Late stage Parkinson’s ($33$ slides), controls ($3$ slides) and iLBD ($8$ slides) cases were selected from Oxford Brain Bank  and the Parkinson’s UK Brain Bank. The control cases, which lack alpha-synuclein pathology, were included as negative controls. \\
We focused on the midbrain area sampled coronally at the level of the substantia nigra, the red nucleus and the oculomotor nerve. Paraffin-embedded blocks were cut into thin sections of $6$ $\mu$m on a microtome for immunohistochemistry. Brightfield immunohistochemistry experiments were performed using antibodies against microglia with the following dilutions: rabbit (019-19741, Wako, Cambridge Biosciences, UK) anti-IBA1 at $1:1000$ and anti-alpha synuclein C110-115 epitope at $1:10000$ \cite{altay2022prominent}. The first step was deparaffinization of formalin-fixed paraffin embedded sections in $100\%$ xylene solution and rehydration in descending concentrations of diluted ethanol ($100\%$, $96\%$, $90\%$, $70\%$). Antigen retrieval was done by heat induced epitope opening using citric acid buffer (pH = $6.2$) for $30$ min in a microwave. Thereafter, sections were pre-treated with dual enzyme block to block endogenous peroxidase and phosphatase activity. Sections were blocked with a solution of $5\%$ Bovine serum albumin + Tween20 ($0.1\%$) + normal horse serum ($5\%$) in 1X PBS and then incubated with primary antibodies overnight. The next day, secondary antibodies were applied using either the Immunopress duet kit (MP7714, Vector labs, UK) with anti-mouse epitopes visualised in magenta with alkaline-phosphatase and anti-rabbit epitopes visualised in brown with DAB, visualizing alpha-synuclein in magenta and microglia in brown. Sections were counterstained with haematoxylin and coverslipped with permanent mounting medium before imaging. Imaging was done using high-resolution histological slide scanners: Aperio Imagescope (Oxford, UK) for analysis at $0.45$ $\mu$m $\times$ $0.45$ $\mu$m pixel resolution.
To be more easily processed, these slides are subdivised in tiles of size $1024 \times 1024$ pixels. 
%%%%%%%%%%%%%%%%%%%%%%%%%%%%%%%%%%%%%%%%%%%%%55
%%%%%%%%%%%%%%%%%%%%%%%%%%%%%%%%%%%%%%%%%%%55
\vspace{-2mm}
\section{Segmentation and classification of alpha-synuclein aggregates}
\vspace{-1mm}
%%%%%%%%%%%%%%%%%%%%%%%%%%%%%%%%%%%%%%
\subsection{Segmentation of alpha-synuclein aggregates}
\vspace{-1mm}
%%%%%%%%%%%%%%%%%%%%%%%%%%%%%%%%%%%%%%
\subsubsection{Stain separation-based segmentation}
\vspace{-1mm}
%%%%%%%%%%%%%%%%%%%%%%%%%%%%%%%%%%%%%%
Our RGB histopathology tiles contain alpha-synuclein stained in magenta via the alkaline-phosphatase substrate. To isolate this signal, we applied the Vahadane stain normalization method~\cite{vahadane2016structure}, which models each pixel’s optical density (OD) as a linear combination of stain-specific basis vectors and decomposes the OD image using Sparse Non-negative Matrix Factorization (SNMF) into a matrix \( \V{W} \) that represents the stain color basis (columns corresponding to hematoxylin, alkaline-phosphatase and DAB vectors), and a matrix \( \V{H} \) that represents the stain concentration maps. \\
The alkaline-phosphatase concentration map, $\mathbf{H}_{\text{alkaline}}$, was then extracted and normalized to $[0,1]$, providing a pixel-wise probability of belonging to an alpha-synuclein aggregate.  A binary mask was obtained by thresholding at $0.5$: $\V{S}_{\text{alkaline}} = \mathbf{H}_{\text{alkaline}} > 0.5$. While this procedure effectively locates aggregates, further refinement is needed for precise segmentation suitable for shape analysis.
%%%%%%%%%%%%%%%%%%%%%%%%%%%%%%%%%%%%%%
\vspace{-1mm}
\subsubsection{Attention-based segmentation} \label{sec:attention_map}
\vspace{-1mm}
%%%%%%%%%%%%%%%%%%%%%%%%%%%%%%%%%%%%%%
Instead of using morphological operations on coarse segmentation $\V{S}_{\text{alkaline}}$ as in \cite{perochon2025unraveling}, which require carefully tuned hyperparameters that do not generalize well across the highly variable shapes and sizes of alpha-synuclein aggregates, we exploit the self-attention mechanism of a Vision Transformer (ViT) \cite{dosovitskiy2020image}. This captures long-range similarities and redundant color and texture patterns, effectively highlighting potential aggregate regions. A classifier composed of a ViT backbone and a linear classification head is trained to detect whether a patch contains an alpha-synuclein aggregate of any shape or size, and the attention maps of its class token provide spatial information about the locations of these aggregates. To train this classifier, we used a DINOv3 backbone \cite{simeoni2025DINOv3}, chosen for its favorable balance between performance and ease of use, with unfrozen weights. The model was trained on $390$ carefully selected tiles, $200$ containing aggregates and $190$ without, split into $292$ training and $98$ validation tiles. Further details regarding the training procedure and hyperparameters are provided in subsection \ref{sec:segmentation}. \\
Using this model, a tile is defined by $\mathbf{x} \in \mathbb{R}^{H \times W \times 3}$,  which is tokenized into $N$ non-overlapping patches (called tokens) and embedded as  $\{\mathbf{z}_i^0\}_{i=1}^{N}$, along with a learnable class token $\mathbf{z}_{\text{cls}}^0$.  At transformer layer $l$, the multi-head self-attention (MHSA) mechanism produces an attention matrix 
$\mathbf{A}^{(l)} \in \mathbb{R}^{h \times (N+1) \times (N+1)}$,  where $h$ is the number of attention heads and $(N+1)$ accounts for the class token.  We are particularly interested in the attention from the \textit{class token} to the patch tokens in the last layer $L$, which can be written as $\mathbf{a}_{\text{cls}} = \frac{1}{h} \sum_{k=1}^{h}  \mathbf{A}^{(L)}_k[\text{cls}, 1:N],$ where $\mathbf{A}^{(L)}_k[\text{cls}, 1:N]$ are the attention weights from the class token to all $N$ patch tokens for head $k$.  
After normalization, the class attention vector is resized to the original image dimensions, and re-normalized to produce a probabilistic map highlighting likely alpha-synuclein aggregate locations. \\
A preliminary mask $\V{S}_{\text{attention}}$ is obtained by thresholding $P$ at $\tau = 0.1$, deliberately set very low to be inclusive, and refined with a fully-connected CRF \cite{krahenbuhl2011efficient} to produce $\V{S}_{\text{refined}}$. Only connected components overlapping with the alkaline-phosphatase stain $\V{S}_{\text{alkaline}}$ are retained, yielding $\V{S}_{\text{combined}}$. Post-processing removes small objects below $T_s$, performs instance segmentation via connected components, and associates components within distance $T_d$ to consolidate fragmented aggregates (see Figure \ref{fig:segmentation_pipeline}). Objects with maximum Feret diameter below $T_F$ are filtered out to remove small, unreliable detections while retaining thin aggregates like neurites, producing the final segmentation mask $\V{S}$.
The hyperparameters involved and the evaluation of the resulting segmentation on the validation dataset are reported in section \ref{sec:experiments}. The final segmentation pipeline is illustrated in Figure \ref{fig:segmentation_pipeline}. We applied the segmentation procedure to all WSI, focusing on the tiles classified by our algorithm as containing aggregates. From these, we constructed an alpha-synuclein aggregate dataset of $4819$ images by extracting a $256 \times 256$ patch centered on the centroid of each aggregate.  These patches were subsequently used as input for the classification stage of the pipeline, described below.
%%%%%%%%%%%%%%%%%%%%%%%%%%%%%%%%%%%%%%
\vspace{-1mm}
\subsection{Image Retrieval and Classification of alpha-synucleins aggregates}
\begin{figure*}[!t]
\centerline{\includegraphics[width=1.8\columnwidth]{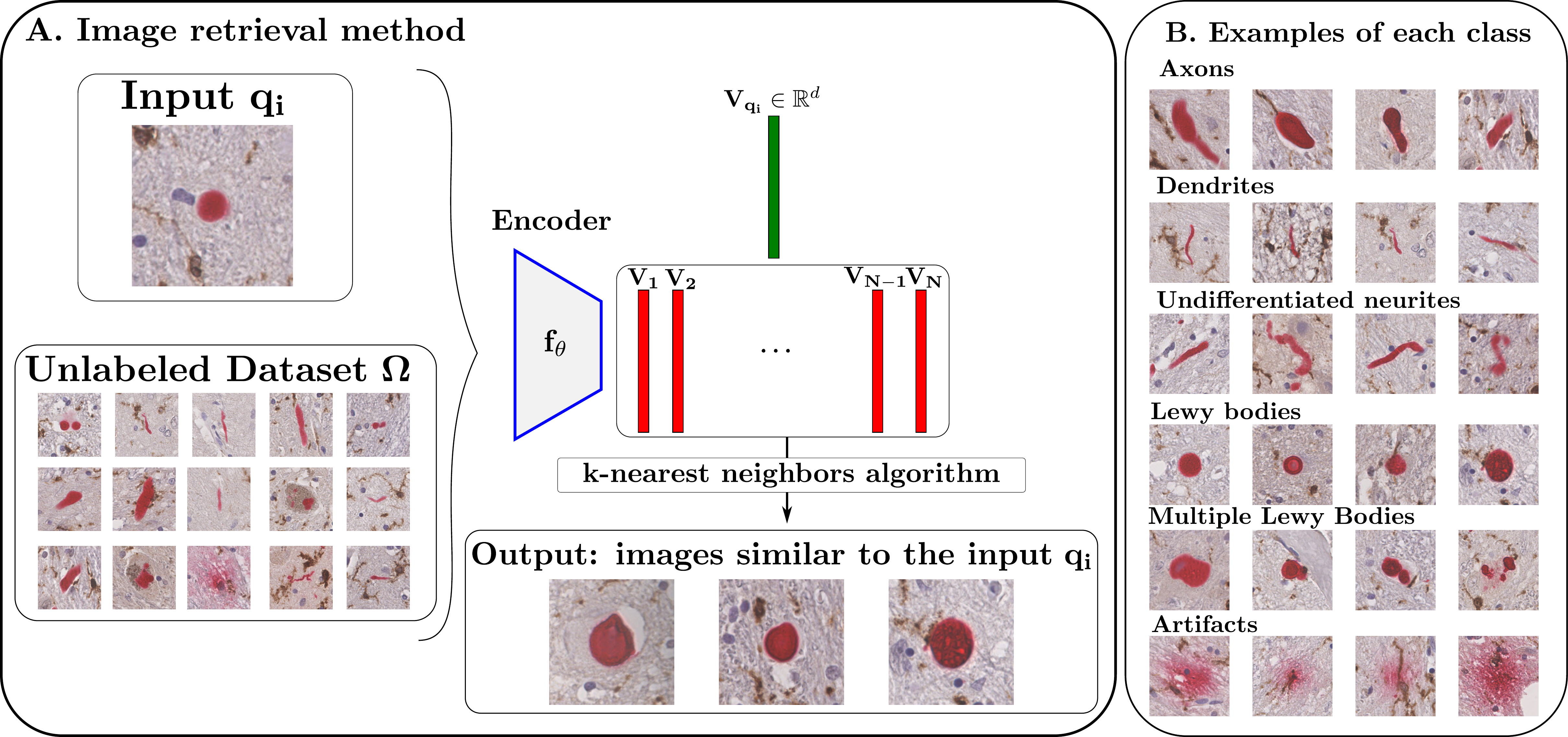}}
\caption{{\bf Image retrieval procedure}. {\bf (A)} Retrieval of images from the unlabeled dataset $\Omega$ that are most similar to the query input image $q_i$, using a $k$-nearest neighbors search on embeddings generated by the neural network $f_{\theta}$. {\bf (B)} Representative examples of alpha-synuclein aggregates for each class, extracted using the described method.}
\label{fig:image_retrieval}
\end{figure*}

\vspace{-1mm}
%%%%%%%%%%%%%%%%%%%%%%%%%%%%%%%%%%%%%%
To classify the aggregates, we first visually identified main classes and characterized their morphology. Lewy bodies are characterized by approximately spherical morphologies. Neurites, in contrast, exhibit elongated morphologies and can correspond to axons, which are typically cylindrical, or dendrites, which are slender. Distinguishing thin axons from thick dendrites, which exhibit similar morphologies and are both neurites, is challenging for pathologists, leading to an “undifferentiated neurites” class. Multiple Lewy bodies may appear closely spaced and are associated in previous instance segmentation, forming clusters of particular pathological interest. In previous instance segmentation, such closely spaced bodies were segmented as a single aggregate through label association of connected components. Finally, some staining artifacts are characterized by irregular and dispersed signal patterns and need to be excluded in subsequent analysis. Based on these considerations, we defined six main classes for analysis: Lewy bodies, axons, dendrites, undifferentiated neurites, multiple Lewy bodies, and artifacts.\\
To construct our dataset, we aimed to recover representative images from each class from our unlabeled set of alpha-synuclein aggregate patches. The goal was to obtain a meaningful representation of the images in a feature space, followed by a nearest neighbors approach to select prototypical images for each class, as illustrated in Figure \ref{fig:image_retrieval}. 
To achieve this, we fine-tuned a DINOv3 backbone with the SimCLR algorithm \cite{chen2020simple} on our unlabeled dataset, to learn feature embeddings in an unsupervised manner (more details in about the training can be found section \ref{sec:experiments}). Next, we chose one image representative of each of the six classes described above as a query and for each of them applied a nearest neighbors search in the feature space to identify the $250$ closest images from the unlabeled dataset. 

% SimCLR trains an encoder to produce embeddings in which augmented views of the same image are close together, while embeddings of different images are far apart. 
% For each image, two augmented views are generated using random crops, flips, color jittering, Gaussian blur, or other deformations.  
% A contrastive loss encourages similarity between positive pairs (augmented views of the same image) and dissimilarity with all other images in the batch. This approach yields meaningful embeddings of the alpha-synuclein aggregate images, capturing visual similarities without requiring manual labels. 

%%%%%%%%%%%%%%%%%%%%%%%%%%%%
%%%%%%%%%%%%%%%%%%%%%%%%%%%%%%%%%%%%%%%
These images were subsequently labeled by an expert neuropathologist. 
After removing potential duplicates and outliers, the resulting dataset consisted of $953$ images composed of $345$ axons, $254$ Lewy Bodies, $119$ dendrites, $83$ undifferentiated neurites, $117$ multiple Lewy Bodies and $35$ staining artifacts. Some representative examples of each classes are then displayed in Figure \ref{fig:image_retrieval}-(B). These images were split into a training set of $664$ images and a validation set of $289$ images. We then trained different classifiers on this dataset, whose architecture and performances are discussed in section \ref{sec:experiments}.
\vspace{-4mm}
%%%%%%%%%%%%%%%%%%%%%%%%%%%%%%%%%%%%%%%%%%%%%%%%%%%%%%%%%%%%%5
\section{Experiments and Results} \label{sec:experiments}
\vspace{-1mm}
\subsection{Segmentation of alpha-synuclein aggregates}\label{sec:segmentation}
\vspace{-1mm}
%%%%%%%%%%%%%%%%%%%%%%%%%%%%%%%%%%%%%%%%%%%%%%%%%%%%%%%%%%%%%
For the tile classification network, given the relative simplicity of the task, we used the AdamW optimizer with a learning rate of $1 \times 10^{-5}$ and a step-based learning rate scheduler.  Training was performed for $20$ epochs using a binary cross-entropy loss.  The checkpoint with the lowest validation loss was selected, and the model achieved a balanced accuracy of $96\%$ on the validation split.\\
We used the same validation dataset for evaluating the segmentation. We used an arbitrary surface threshold of $T_s = 100$ square pixels to remove noisy objects, and empirically determined the associating distance threshold $T_d = 20$ pixels and the maximum Feret diameter threshold $T_F = 33$ pixels through a grid-search procedure. We used the library \texttt{SimpleCRF} \cite{SimpleCRF} with the default hyperparameters.
To reduce annotation workload, we used a partial annotation strategy similar to \cite{perochon2025unraveling}: an expert neuropathologist manually counted alpha-synuclein aggregates for the test set tiles, which were then compared to those detected by our algorithm. Segmentation masks were visually assessed and categorized as Good, Medium, or Bad. Overall, the algorithm showed an average relative difference of $24\%$ compared to manual counts, and among the $58$ segmented aggregates, $90\%$ of masks were rated Good, $7\%$ Medium, and $3\%$ Bad.
%%%%%%%%%%%%%%%%%%%%%%%%%%%%%%%%%%%%%%
\vspace{-3mm}
\subsection{Classification of alpha-synuclein aggregates}
%%%%%%%%%%%%%%%%%%%%%%%%%%%%%%%%%%%%%%
For the image retrieval, the embedding network training used AdamW with batch size $128$, learning rate $2 \times 10^{-4}$ for both backbone and linear head, weight decay $0.05$, $\beta_1=0.9$, $\beta_2=0.95$, and a warmup cosine scheduler. For downstream classification, ResNet-50 (ImageNet pretrained), DINOv3, and CONCH \cite{roodveldt2024immune} were combined with a linear head with a dropout of $0.2$ and ReLU activation. Two strategies were tested: full fine-tuning with AdamW (batch size $32$, weight decay $1 \times 10^{-4}$, learning rate $1 \times 10^{-4}$ with warmup cosine scheduler) and training only the linear head on frozen backbones (learning rate of $1 \times 10^{-3}$ with warmup cosine scheduler). The checkpoint with the lowest validation loss was selected, and balanced accuracies are reported in Table \ref{tab:backbone_performance}.
%%%%%%%%%%%%%%%%%%%%%%%%%%%%%%%%%%%%%%
\begin{table}[h!]
\centering
\caption{Classification performance of different backbones on the alpha-synuclein aggregate dataset. Balanced accuracy (\%) is reported for full fine-tuning (FT) and frozen backbone.}
\begin{tabular}{lcc}
\hline
\textbf{Backbone} & \textbf{Full FT (\%)} & \textbf{Frozen (\%)} \\
\hline
ResNet-50 (ImageNet) & \textbf{80.65} & 16.67 \\
DINOv3 & 79.26 & 69.73 \\
CONCH (histopathology) & 76.76 & 55.18 \\
\hline
\end{tabular}
\label{tab:backbone_performance}
\end{table}
\vspace{-1mm}
\vspace{-1mm}
%%%%%%%%%%%%%%%%%%%%%%%%%%%%%%%%%%%%%%
\section{Discussion}
\vspace{-1mm}
%%%%%%%%%%%%%%%%%%%%%%%%%%%%%%%%%%%%%%
The results in subsection \ref{sec:segmentation} confirm that our tile classifier reliably detects tiles containing alpha-synuclein aggregates. This capability allows the network’s attention layers to be used for segmentation (subsubsection \ref{sec:attention_map}), producing high-quality masks (subsection \ref{sec:segmentation}) suitable for subsequent shape and spatial analyses. Nevertheless, a relative error of $24\%$ in aggregate detection indicates substantial room for improvement. For example, the label-association step could be enhanced by incorporating aggregate shapes rather than relying solely on distance thresholds, which would help distinguish fragmented objects from genuinely separate ones. Additionally, since the method cannot reliably separate closely spaced Lewy bodies, all images containing multiple Lewy bodies were grouped into a single class (as shown in Figure \ref{fig:image_retrieval}-(B)), making it difficult to determine the exact number of Lewy bodies per image and potentially affecting the accuracy of quantitative analyses.\\
For the classification of aggregates, a comparatively simple network such as a ResNet-50 pretrained on ImageNet outperformed DINOv3, and achieving $80\%$ balanced accuracy is notable given the difficulty of distinguishing some classes. This observation aligns with \cite{liu2025does}, which reported that DINOv3 features do not transfer effectively to histopathology tasks, even though the ResNet-50 must be fully fine-tuned to achieve this high performance. In contrast, DINOv3 benefits from full fine-tuning but still performs reasonably well even with a frozen backbone, suggesting that its pretrained features are more generally representative. Additionally, CONCH, a large foundation model specifically designed for histopathology, did not surpass the performance of DINOv3 on this dataset under either full fine-tuning or frozen backbone conditions. This outcome may be explained by the relatively small size of our dataset and the particular visual characteristics of magenta-stained aggregates. Indeed, the translation-equivariance property of CNNs enables them to generalize from fewer data points, whereas transformers require extremely large datasets to learn similar translation-invariance. While DINOv3 may have already acquired this invariance during pretraining, CONCH may not have, possibly due to a histopathological domain shift.  This demonstrates that a relatively lightweight classifier can outperform larger models trained on massive datasets, achieving competitive performance with lower computational cost, suggesting that current foundation models may not yet be fully generalizable.\\
Despite strong performance, future improvements could come from leveraging unlabeled images with self-supervised or semi-supervised methods \cite{caron2020unsupervised, sohn2020fixmatch}, and including less common classes such as astrocytic aggregates \cite{altay2022prominent} or closely spaced neurites.
%%%%%%%%%%%%%%%%%%%%%%%%%%%%%%%
\vspace{-1mm}
\vspace{-1mm}
\section{Conclusion}
\vspace{-1mm}
%%%%%%%%%%%%%%%%%%%%%%%%%%%%%%%
In this study, we developed an automated pipeline for the detection, segmentation, and classification of alpha-synuclein aggregates in midbrain tissue from PD and iLBD cases. Our approach combines a weakly supervised segmentation method robust to staining variability with a neural network classifier capable of distinguishing major aggregate morphologies, and incorporates a self-supervised nearest-neighbors retrieval algorithm that could serve to assist neuropathologists for annotation and classification. The resulting segmentation masks are of sufficient quality for downstream shape and spatial analysis. For classification, a relatively simple ResNet-50 backbone outperformed larger foundation models such as DINOv3 and CONCH, demonstrating the effectiveness of lightweight architectures on histopathology data. The pipeline enables large-scale morphological analysis of alpha-synuclein aggregates, and their spatial interactions with microglial phenotypes.\\
Potential improvements include improved separation of closely spaced Lewy bodies, refined label association strategies, and leveraging unlabeled data through self- or semi-supervised learning. Expanding the classification dataset to include additional and rare aggregate classes could further enhance performance. 
%%%%%%%%%%%%%%%%%%%%%%%%%%%%%%%%
\section{Compliance with ethical standards}
%%%%%%%%%%%%%%%%%%%%%%%%%%%%%%%
The study was conducted according to the guidelines of the Declaration of Helsinki and approved by the ethics committees at the Oxford Brain Bank (Rec approval: 23/sc/0241, South Central Oxford C) and the Parkinson’s UK Brain Bank.
%%%%%%%%%%%%%%%%%%%%%%%%%%%%%%%5
\section{Acknowledgments}
DAM was funded by a Springboard grant funded by the British Council (grant agreement No 1170803491). D. H. group is funded by ANR AstroXcite  and the European Research Council (ERC) under the European Union’s Horizon 2020 research and innovation program (No 882673).
%%%%%%%%%%%%%%%%%%%%%%%%%%%%%%%%%%%%%%%%%%%5
\bibliographystyle{IEEEbib}
\bibliography{strings,refs}

\begin{thebibliography}{10}

\bibitem{dauer2003parkinson}
William Dauer and Serge Przedborski,
\newblock ``{P}arkinson's disease: mechanisms and models,''
\newblock {\em Neuron}, vol. 39, no. 6, pp. 889--909, 2003.

\bibitem{cheng2010clinical}
Hsiao-Chun Cheng, Christina~M Ulane, and Robert~E Burke,
\newblock ``Clinical progression in {P}arkinson disease and the neurobiology of axons,''
\newblock {\em Annals of Neurology}, vol. 67, no. 6, pp. 715--725, 2010.

\bibitem{altay2022prominent}
Melek~Firat Altay, Alan King~Lun Liu, Janice~L Holton, Laura Parkkinen, and Hilal~A Lashuel,
\newblock ``Prominent astrocytic alpha-synuclein pathology with unique post-translational modification signatures unveiled across {L}ewy body disorders,''
\newblock {\em Acta Neuropathologica Communications}, vol. 10, no. 1, pp. 163, 2022.

\bibitem{dijkstra2014stage}
Anke~A Dijkstra, Pieter Voorn, Henk~W Berendse, Henk~J Groenewegen, Netherlands~Brain Bank, Annemieke~JM Rozemuller, and Wilma~DJ van~de Berg,
\newblock ``Stage-dependent nigral neuronal loss in incidental {L}ewy body and {P}arkinson's disease,''
\newblock {\em Movement Disorders}, vol. 29, no. 10, pp. 1244--1251, 2014.

\bibitem{prasad2012critical}
Kavita Prasad, Thomas~G Beach, John Hedreen, and Eric~K Richfield,
\newblock ``Critical role of truncated $\alpha$-synuclein and aggregates in {P}arkinson's disease and incidental {L}ewy body disease,''
\newblock {\em Brain Pathology}, vol. 22, no. 6, pp. 811--825, 2012.

\bibitem{roodveldt2024immune}
Cintia Roodveldt, Liliana Bernardino, Ozgur Oztop-Cakmak, Milorad Dragic, Kari~E Fladmark, Sibel Ertan, Busra Aktas, Carlos Pita, Lucia Ciglar, Gaetan Garraux, et~al.,
\newblock ``The immune system in {P}arkinson's disease: what we know so far,''
\newblock {\em Brain}, vol. 147, no. 10, pp. 3306--3324, 2024.

\bibitem{andersen2025sympathetic}
Katrine~B Andersen, Anushree Krishnamurthy, Mie~Kristine Just, Nathalie Van Den~Berge, Casper Skj{\ae}rb{\ae}k, Jacob Horsager, Karoline Knudsen, Jacob~W Vogel, Jon~B Toledo, Johannes Attems, et~al.,
\newblock ``Sympathetic and parasympathetic subtypes of body-first {L}ewy body disease observed in postmortem tissue from prediagnostic individuals,''
\newblock {\em Nature Neuroscience}, pp. 1--12, 2025.

\bibitem{perochon2025unraveling}
Theo Perochon, Zeljka Krsnik, Marco Massimo, Yana Ruchiy, Alejandro~Lastra Romero, Elyas Mohammadi, Xiaofei Li, Katherine~R Long, Laura Parkkinen, Klas Blomgren, et~al.,
\newblock ``Unraveling microglial spatial organization in the developing human brain with deepcellmap, a deep learning approach coupled with spatial statistics,''
\newblock {\em Nature Communications}, vol. 16, no. 1, pp. 1577, 2025.

\bibitem{vatsa2024network}
Naman Vatsa, Julia~K Brynildsen, Thomas~M Goralski, Kevin Kurgat, Lindsay Meyerdirk, Libby Breton, Daniella DeWeerd, Laura Brasseur, Lisa Turner, Katelyn Becker, et~al.,
\newblock ``Network analysis of $\alpha$-synuclein pathology progression reveals p21-activated kinases as regulators of vulnerability,''
\newblock {\em bioRxiv}, 2024.

\bibitem{pearce2022automatic}
Bradley Pearce, Peter Coetzee, Duncan Rowland, Scott Linfoot, David~T Dexter, Djordje Gveric, and Stephen Gentleman,
\newblock ``Automatic sample segmentation \& detection of {P}arkinson’s disease using synthetic staining \& deep learning,''
\newblock {\em bioRxiv}, pp. 2022--08, 2022.

\bibitem{barber2025development}
A~Barber-Janer, E~Van~Acker, E~Vonck, D~Plessers, F~Rosada, C~Van~den Haute, V~Baekelandt, and W~Peelaerts,
\newblock ``Development of convolutional neural networks for automated brain-wide histopathological analysis in mouse models of synucleinopathies,''
\newblock {\em bioRxiv}, pp. 2025--07, 2025.

\bibitem{dadgar2022mesoscale}
Ehsan Dadgar-Kiani, Gregor Bieri, Ronald Melki, Aaron~D Gitler, and Jin~Hyung Lee,
\newblock ``Mesoscale connections and gene expression empower whole-brain modeling of $\alpha$-synuclein spread, aggregation, and decay dynamics,''
\newblock {\em Cell Reports}, vol. 41, no. 6, 2022.

\bibitem{vahadane2016structure}
Abhishek Vahadane, Tingying Peng, Amit Sethi, Shadi Albarqouni, Lichao Wang, Maximilian Baust, Katja Steiger, Anna~Melissa Schlitter, Irene Esposito, and Nassir Navab,
\newblock ``Structure-preserving color normalization and sparse stain separation for histological images,''
\newblock {\em IEEE Transactions on Medical Imaging}, vol. 35, no. 8, pp. 1962--1971, 2016.

\bibitem{dosovitskiy2020image}
Alexey Dosovitskiy,
\newblock ``An image is worth 16x16 words: Transformers for image recognition at scale,''
\newblock {\em arXiv preprint arXiv:2010.11929}, 2020.

\bibitem{simeoni2025DINOv3}
Oriane Sim{\'e}oni, Huy~V Vo, Maximilian Seitzer, Federico Baldassarre, Maxime Oquab, Cijo Jose, Vasil Khalidov, Marc Szafraniec, Seungeun Yi, Micha{\"e}l Ramamonjisoa, et~al.,
\newblock ``{DINOv3},''
\newblock {\em arXiv preprint arXiv:2508.10104}, 2025.

\bibitem{krahenbuhl2011efficient}
Philipp Kr{\"a}henb{\"u}hl and Vladlen Koltun,
\newblock ``Efficient inference in fully connected crfs with gaussian edge potentials,''
\newblock {\em Advances in Neural Information Processing Systems}, vol. 24, 2011.

\bibitem{chen2020simple}
Ting Chen, Simon Kornblith, Mohammad Norouzi, and Geoffrey Hinton,
\newblock ``A simple framework for contrastive learning of visual representations,''
\newblock in {\em International Conference on Machine Learning}. PmLR, 2020, pp. 1597--1607.

\bibitem{SimpleCRF}
Healthcare~Intelligence Laboratory,
\newblock ``Simple{CRF}: Matlab and python wrap of conditional random field ({CRF}) and fully connected (dense) {CRF} for 2{D} and 3{D} image segmentation,'' .

\bibitem{liu2025does}
Che Liu, Yinda Chen, Haoyuan Shi, Jinpeng Lu, Bailiang Jian, Jiazhen Pan, Linghan Cai, Jiayi Wang, Yundi Zhang, Jun Li, et~al.,
\newblock ``Does {DINOv3} set a new medical vision standard?,''
\newblock {\em arXiv preprint arXiv:2509.06467}, 2025.

\bibitem{caron2020unsupervised}
Mathilde Caron, Ishan Misra, Julien Mairal, Priya Goyal, Piotr Bojanowski, and Armand Joulin,
\newblock ``Unsupervised learning of visual features by contrasting cluster assignments,''
\newblock {\em Advances in Neural Information Processing Systems}, vol. 33, pp. 9912--9924, 2020.

\bibitem{sohn2020fixmatch}
Kihyuk Sohn, David Berthelot, Nicholas Carlini, Zizhao Zhang, Han Zhang, Colin~A Raffel, Ekin~Dogus Cubuk, Alexey Kurakin, and Chun-Liang Li,
\newblock ``Fixmatch: Simplifying semi-supervised learning with consistency and confidence,''
\newblock {\em Advances in Neural Information Processing Systems}, vol. 33, pp. 596--608, 2020.

\end{thebibliography}
\end{document}